# The Observatory for Multi-Epoch Gravitational Lens Astrophysics (OMEGA)

Leonidas A Moustakas[*,a], Adam J Bolton[b], Jeffrey T Booth[a], James S Bullock[c],
Edward Cheng[d], Dan Coe[a], Christopher D Fassnacht[e], Varoujan Gorjian[a], Cate Heneghan[a],
Charles R Keeton[f], Christopher S Kochanek[g], Charles R Lawrence[a], Philip J Marshall[h],
R Benton Metcalf[i], Priyamvada Natarajan[j], Shouleh Nikzad[a], Bradley M Peterson[g],
Joachim Wambsganss[k]
for the OMEGA Team

[a]Jet Propulsion Laboratory, Caltech, MS 169-327, 4800 Oak Grove Dr, Pasadena, CA 91109;
[b]Institute for Astronomy, University of Hawaii at Manoa, 2680 Woodlawn Dr, Honolulu, HI 96822;
[c]Physics & Astronomy Department, University of California, Irvine, CA 92697-4575;
[d]Conceptual Analytics, LLC, 8209 Woburn Abbey Road, Glenn Dale, MD 20769;
[e]Physics Department, One Shields Ave, University of California, Davis, CA 95616;
[f]Dept of Physics & Astronomy, Rutgers University, 136 Frelinghuysen Road, Piscataway, NJ 08854;
[g]Dept of Astronomy, The Ohio State University, 140 W 18[th] Ave, Columbus, OH 43210-1173;
[h]Physics department, University of California, Santa Barbara, CA 93106;
[i]Max Planck Institute for Astrophysics, Karl-Schwarzschild-Str 1, 85741 Garching, Germany;
[j]Dept of Astronomy, Yale University, 260 Whitney Ave, New Haven, CT 06511;I ho
[k]Astronomisches Rechen-Institut Moenchhofstr 12-14 69120 Heidelberg, Germany

**ABSTRACT**

Dark matter in a universe dominated by a cosmological constant seeds the formation of structure and is the scaffolding for galaxy formation. The nature of dark matter remains one of the fundamental unsolved problems in astrophysics and physics even though it represents 85% of the mass in the universe, and nearly one quarter of its total mass-energy budget. The mass function of dark matter "substructure" on sub-galactic scales may be enormously sensitive to the mass and properties of the dark matter particle. On astrophysical scales, especially at cosmological distances, dark matter substructure may only be detected through its gravitational influence on light from distant varying sources. Specifically, these are largely active galactic nuclei (AGN), which are accreting super-massive black holes in the centers of galaxies, some of the most extreme objects ever found. With enough measurements of the flux from AGN at different wavelengths, and their variability over time, the detailed structure around AGN, and even the mass of the super-massive black hole can be measured. The Observatory for Multi-Epoch Gravitational Lens Astrophysics (OMEGA) is a mission concept for a 1.5-m near-UV through near-IR space observatory that will be dedicated to frequent imaging and spectroscopic monitoring of ~100 multiply-imaged active galactic nuclei over the whole sky. Using wavelength-tailored dichroics with extremely high transmittance, efficient imaging in six channels will be done simultaneously during each visit to each target. The separate spectroscopic mode, engaged through a flip-in mirror, uses an image slicer spectrograph. After a period of many visits to all targets, the resulting multidimensional movies can then be analyzed to a) measure the mass function of dark matter substructure; b) measure precise masses of the accreting black holes as well as the structure of their accretion disks and their environments over several decades of physical scale; and c) measure a combination of Hubble's local expansion constant and cosmological distances to unprecedented precision. We present the novel OMEGA instrumentation suite, and how its integrated design is ideal for opening the time domain of known cosmologically-distant variable sources, to achieve the stated scientific goals.

**Keywords:** Strong gravitational lensing, time domain, dark matter, expansion history, active galactic nuclei, space telescope, spectroscopy, astronomical survey

---

[*] leonidas@jpl.nasa.gov; phone 1 818 393 5095; fax 818 354 1004; http://www.its.caltech.edu/~leonidas

*Everything flows, nothing stands still (Heraclitus)*

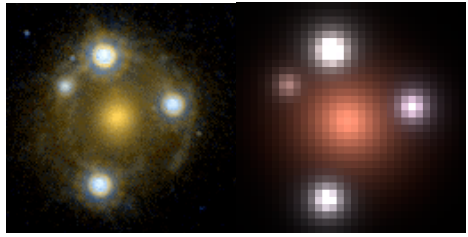

Figure 1. SDSS J0924+0219: A very luminous AGN (a quasar), gravitationally lensed into four images by a foreground massive elliptical galaxy[1]. The brightnesses and colors of the images vary with time through a combination of astrophysical effects. The left image shows a *Hubble* Space Telescope Advanced Camera for Surveys image (~3.5x3.5arcsec field), and the right image (using a different color-combination) shows a realization of how this lens would appear at the resolution and sampling of OMEGA.

## 1. INTRODUCTION

### 1.1. The Observatory for Multi-Epoch Gravitational Lens Astrophysics: An executive summary

**OMEGA's goals.** The Observatory for Multi-Epoch Gravitational-Lens Astrophysics (OMEGA) mission concept is a 1.5 m telescope diffraction-limited at 650 nm, with photometric imager and spectroscopic image slicer instruments, in a halo orbit at the second Earth-Sun Lagrange point (L2). The primary science goals OMEGA is designed for are 1) to determine the *nature of the dark matter* that fills the Universe; 2) to measure *precise black hole masses*, and the structure of *black hole accretion disks* and their environments; and 3) to make geometric measurements of *Hubble's constant and cosmological distances* with unprecedented precision.

**Mission summary.** OMEGA will monitor ~100 previously-identified gravitationally-lensed active galactic nuclei (AGN) in six broad-band filters spanning the near-ultraviolet (NUV) to the near-infrared (NIR) for high-precision photometry at a fast cadence (hours to days), and with moderate-resolution integral-field spectroscopy on a slower cadence (days to weeks). OMEGA will efficiently observe targets over the full sky with minimal overheads and observing gaps. At the end of its five-year mission, OMEGA will have obtained well-sampled "movies" of the time variability of the full sample as a function of wavelength (in both broadband imaging and spectroscopic channels), ultra-deep images of each 5.2x5.2 arc-minute field, and ultra-deep two-dimensional spectroscopy of the lensing galaxy. OMEGA will open the time domain in high spatial resolution imaging and spectroscopy of variable strong gravitational lenses with a powerful payload that is based on tried and true technology, but implemented in a novel way.

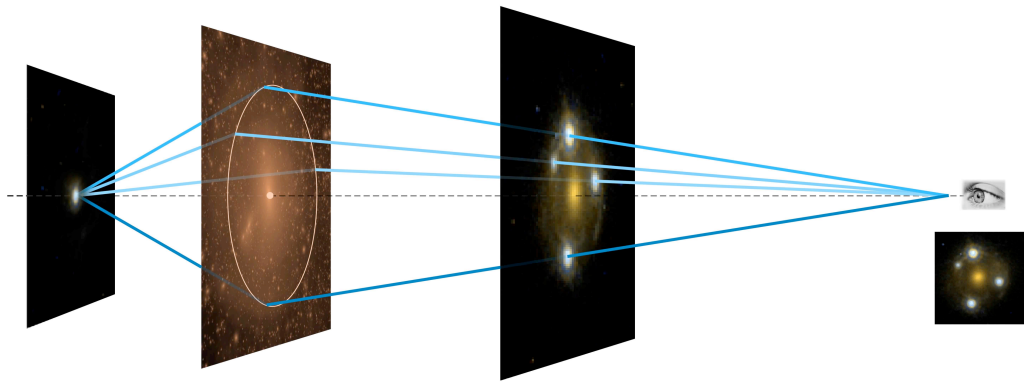

Figure 2. A gravitational lens diagram. Light from the quasar (left) is bent and focused by the massive lens galaxy (center), into multiple images (right). Each ray shown has a different path, and so travel time. The dark matter substructure, as well as the stars in the lens galaxy, all influence the image magnifications. The intensity variations along each ray, each of which forms an image, illustrate the possible variability with time.

## 1.2. OMEGA's Targets: Gravitationally-Lensed Active Galactic Nuclei

**Gravitational lenses.** To frame the OMEGA mission and the objectives driving its design, we begin with a brief review of the physical properties of its intended targets. These are multiply-imaged, gravitationally lensed AGN. An example of one such object is shown in Figures 1 and 2. There is a galaxy in the center of the image. The four point sources surrounding it are each an image of a single luminous AGN, which by good fortune is very closely aligned with the line connecting us and the galaxy. That direct view is obscured by the galaxy itself; but its gravity is great enough that it has focused light from the source, as illustrated in Figure 2, to produce four images of the background source[3]. It is immediately apparent from Figure 1 that the brightness of the four images is *not* the same as may naively be expected. Therein lay the subtle advantages we may take from this spectacular phenomenon.

**Accreting super-massive black holes.** OMEGA's targets, AGN, are accreting super-massive black holes with masses $M_{BH}=10^8 M_8$ solar masses ($M_{sun}$), which are hosted in otherwise-normal galaxies[2]. (And indeed, the host galaxy of the AGN in Figure 1 can be seen gravitationally lensed into an "Einstein ring" encircling the lens galaxy.) For the most luminous AGN, *quasars*, $M_8$ is of order unity. Their luminosity is emitted by gaseous material gravitationally accreting through an *accretion disk* with an inner edge of approximately $10^{14} M_8$ cm. Beyond this inner edge, the disk temperature $T$ declines roughly as $T \propto R^{-3/4}$, so the effective size of the emission region depends on wavelength roughly as $R_\lambda \approx 2\times10^{15} \mathrm{x} (\lambda/100\mathrm{nm})^{4/3} \mathrm{x} M_8^{2/3}$ cm. In other words, the *effective size of an AGN disk depends on the wavelength used to observe it*, and a large range of wavelengths will therefore probe different physical scales of the accretion disk[4]. This fact leads to the power a multi-wavelength space mission, and is exploited thoroughly by OMEGA's design, as will be described.

**The advantage of redshift.** The temperature profile of the disk peaks at the inner edge, which corresponds to a (rest-frame) wavelength of $\lambda_{in} \approx 100 M_8^{1/4}$ nm. Photometry at such rest-ultraviolet wavelengths will therefore be sensitive to the inner edge of the accretion disk, which is also the most highly variable (in both amplitude and time) part of an AGN. Since OMEGA's AGN targets are at cosmological distances with redshift $z$, rest-frame wavelengths will be observed to be stretched to redder wavelengths by a factor of $(1+z)$, tempering how far into the ultraviolet it is necessary to go to "see" the inner edge of the accretion disk.

**Larger scale AGN structure.** As it travels outwards from the vicinity of the black hole, the emission from the accretion disk pumps the extended emission line regions, which include the *broad line region* on scales of $R_{CIV} \approx 6\times10^{17} M_8^{1/2}$ cm (for the CIV line), and the much larger *narrow line region* (10 - 100 parsecs, or $\sim 3\mathrm{x}10^{19}$ - $3\mathrm{x}10^{20}$cm). When the accretion rate varies, the continuum luminosity changes, and the emission lines eventually respond to the change as the signal propagates outwards from the accretion disk.

**Scales from lensing.** The gravitational field of an object of mass $M_{lens}$ solar masses deflects light rays by $\theta_{lens} \approx M_{lens}^{1/2}$ µas (micro-arcseconds), which corresponds to a length scale at the source of order $R_{lens} \approx 10^{16} M_{lens}^{1/2}$ cm. Thus, a typical galaxy with $M_{lens} \approx 10^{12} M_{sun}$ produces multiple images with a characteristic separation of order an arcsecond (as seen in Figure 1). We cannot directly image the effects of massive "sub-structure"[5] (such as dwarf galaxy satellites) with $M_{lens} \approx 10^6 M_{sun}$, or stars with $M_{lens} \approx M_{sun}$, but we can detect them through their effects on the brightness of the observed images and (for some mass scales) through observable changes in the light-travel times. Importantly, the substructure effects depend on the size of the *source* (i.e. the accretion disk as it appears at a given wavelength), with small effects for components large compared to $R_{lens}$ and large effects for source components small compared to $R_{lens}$.

**Variability in lensed AGN.** We see time variability in the brightness of lensed images due to multiple phenomena. First, AGN are intrinsically variable. Quasars vary typically by ~20% per year, and lower-luminosity AGN may vary a great deal more. Within each image's light curve, these **continuum flux variations** appear later in the broad emission lines, with a time lag τ corresponding to the light travel time from the accretion disk to the broad line region. For the speed of light "c," this lag is $\tau \approx R_{CIV}/c$, which is on the order of months. Measuring τ determines the size of the broad line region and, through local calibrations, $M_{BH}$ to $\approx 30\%$[2]. This is known as **emission line reverberation mapping**, and is only possible from the ground for AGN that are (cosmologically speaking) in our very neighborhood. In a lens, it takes a different amount of time for light to reach us for each image, so any variation in the source brightness appears in the lensed images with a relative time delay of days to years. These **gravitational time delays**, measured by cross correlating the light curves for different images, are related to the distance to the lens and the distribution of matter in the lens. We also observe uncorrelated, wavelength-dependent brightness changes in each image due to **microlensing** by the stars in the lens galaxy. The scale $R_{lens}$ of the stars is small enough that the magnification of the accretion disk by the stars changes on time scales between weeks and years, leading to variability. The scale $R_{lens}$ is also comparable to the size of the accretion disk $R_\lambda$ so the amplitude of the microlensing variability will depend on wavelength (bigger amplitudes for smaller regions or shorter wavelengths). Microlensing has little effect on the emission lines since their physical scale is much larger than $R_{lens}$ for stars.

**OMEGA.** OMEGA is designed to measure as a function of time and wavelength the variability of lenses created by all four of these four phenomena: AGN continuum variability, AGN emission line reverberation, gravitational lens time delays, and microlensing.

## 2. SCIENTIFIC MOTIVATION AND OBJECTIVES

**The nature of dark matter.** The nature of dark matter remains one of the fundamental unsolved problems in astrophysics and physics even though it represents 85% of the material in the universe. Dark matter in a universe dominated by a cosmological constant ("ΛCDM") seeds the formation of structure and is the scaffolding for galaxy formation. To date, ΛCDM predictions of matter density fluctuations successfully match many observations on scales from $\sim 10^4$ Mpc (the horizon scale) to ~1 Mpc (the distances between galaxies). This success does not seem to extend to scales far below ~1 Mpc, motivating the need for precise observational measurements of small-scale dark matter "substructure" ($\sim 10^6$-$10^8 M_{sun}$) within galaxies. Gravitational lenses provide such a tool because they can detect dark matter substructure gravitationally, in a large number of distinct, distant (lensing) galaxies. Figure 3 shows how different dark matter particle masses may lead to different substructure mass function predictions[7]. OMEGA will measure this substructure, and thus constrain dark matter properties on critical small scales, through two complementary methods using the same data: anomalous flux ratios, and time-delay perturbations[6]. Each of these is sensitive to different moments of the mass function of dark matter substructure, if they may be determined with sufficient precision for each target. Time delay precisions approaching ~0.5% will be possible with OMEGA in many cases, making this experiment possible.

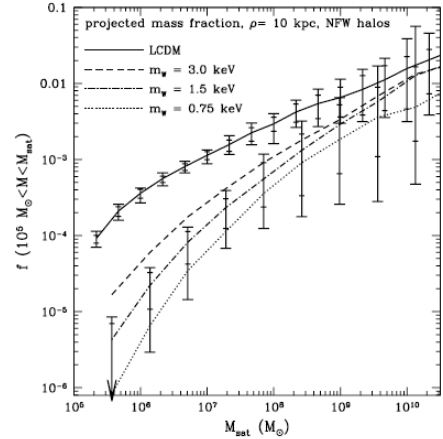

Figure 3. The fraction of dark matter locked in substructure of different mass scales, in a typical massive-galaxy size dark matter halo. This is the critical measurement which will be made to high precision with OMEGA, as a function of lens-galaxy sub-samples that will span galaxy mass, and redshift. Note lines for different assumed dark matter particle masses; and sensitivity thereof especially at the lower mass range!

**Black hole masses and AGN structure.** Although gravitational lensing is intrinsically achromatic, the fact that accretion disks have a definite temperature (and so color) profile leads to wavelength-dependent microlensing by the stars in the lensing galaxy. By observing the microlensing light curves at many wavelengths, the temperature profile of the accretion disk may therefore be reconstructed, from nearly the inner edge of the disk itself. Through the general variations in the continuum light, and the emission lines (from the spectroscopic monitoring combined with the photometric monitoring), full reverberation maps will be built, allowing us to make accurate estimates of black hole masses. At present, microlensing results exist only for 10 systems with disk sizes $R_\lambda$ measured with accuracies of a factor of two[8] and the scaling with wavelength is measured for only one system[9]. Reverberation lags $\tau$ have been measured for about 20 very nearby AGN[10], but only for the emission line as a whole rather than as a function of velocity in the line. The goal of OMEGA is to revolutionize quantitative measurements of accretion disk and line emitting regions and with vastly more detail for each system, both for the intrinsic value of understanding the gravitational dynamics in the vicinity of the densest objects known in the universe, but also for associating their formation and evolution to the corresponding evolution of their host galaxies.

**The scale of the universe: *Hubble*'s constant and cosmography.** The time delays in strong lenses lead directly to a combination of the local expansion rate (*Hubble*'s constant, $H_0$), and ratios of the angular diameter distances that connect the observer, the lens, and the source. OMEGA has the potential to measure $H_0$ to better than 1% precision, a many-fold improvement over what is possible today[11], and a critical ingredient to interpreting future measurements of the expansion history of the universe[12]. This is accomplished "for free," as the time delay precision requirement is driven by the primary science goal, of measuring the mass function of dark matter sub-structure.

## 3. OMEGA MISSION DESIGN

### 3.1 Mission design overview

The science requirements quantified in Table 1 drive the need for space: **1)** Long term photometric precision and stability; **2)** Access to the near-ultraviolet where AGN flux variations are maximal and correspond to very small physical scales; **3)** High angular resolution with a stable point spread function (PSF) over the full near-ultraviolet through near-infrared wavelength range, diffraction-limited around 650nm, for precision photometry of point sources separated by fractions of one arc-second; **4)** Maximal access to most of the sky over the full course of each year; and **5)** Ability to efficiently and rapidly move through a large set of targets with by-object tailored cadences.

We propose a mission with five years of operations in a halo or Lissajous orbit at the Earth-Sun second Lagrange point (L2). The spacecraft consists of a spacecraft bus and an observatory. The observatory is defined to be the 1.5 m primary diameter telescope (including the optics and supporting structure) and two instruments (one imaging system designed to image the *same* field on the sky simultaneously in six different bands, an image slicer channel for integral-field spectroscopy, and associated electronics and structure).

The observing strategy for OMEGA is uniquely tailored to the need for exploring the time domain of lensed AGN. Using a known target list, a preset mission operation plan will schedule observation times for each target. The total integration time for each observation, and the revisit time, or cadence for target, will be tailored to its particular characteristics. Lenses could be visited hourly or every few days (for imaging), and monthly (for spectroscopy). Though complex, the observing strategy will be scheduled in advance, while also allowing the capability to redirect the mission for critical or new events.

Instrument and optical thermal stability required by the precision photometry, and maximum viewing capability on the sky are the primary drivers for the L2 orbit choice. Data rates are relatively low due to the small focal plane sizes, and estimates for a 70% observing efficiency show that OMEGA's data rate will be approximately 100 GB/week. This is not a challenge for downlink from L2 using standard X-band communications. Passive cooling at L2 allows the focal planes to operate at approximately 150K, which meets detector temperature needs.

The General Dynamics SA-200HP RSDO spacecraft bus is baselined for the OMEGA mission. Modeled payload mass (500kg), power (<200W), and propellant (<100m/s) requirements are met by the standard SA-200HP (with a modification of a Fine Guidance Sensor and improved reaction wheels for pointing requirements). Other minor modifications will include a gimbaled telecom antenna, upgraded onboard data storage (from 100Gb to 1.6Tb) to allow a single weekly downlink, and selected subsystem redundancy for the five-year mission duration.

Based on previous Team X studies at JPL of analogous configurations using the SA-200HP bus, we estimate the bus mass to be 475kg. The launch mass of the spacecraft is estimated to be 1330kg with a 30% mass contingency on the observatory and upgrades to the bus and a 15% mass contingency for RSDO bus components. This leaves a 51% launch mass margin on the smallest Delta IV configuration and provides opportunities for launch vehicle sharing.

### 3.1 Telescope

The heart of the observatory is a 1.5m class three-mirror anastigmatic (TMA) telescope that provides the required 0.11arcsec resolution (Figure 4). The telescope is diffraction limited across its 5.2x5.2 arcmin field of view with an f-ratio of 16.3 allowing reasonable PSF sampling with 10.5μm pixels (0.09 arcsec/pixel). Other features incorporated into the design are: polarization insensitivity, achieved with a pair of compensating orthogonal fold mirrors, and using mirror coatings with appropriate dielectric thickness; and a long back focal length, required to accommodate the series of dichroics for the different bands, while accommodating a flip-in mirror for the integral field unit image slicer (IFS).

Table 1: The science goals driving the measurement and mission requirements for OMEGA.

| OMEGA Mission Science Traceability Matrix | Parameter | The Nature of Dark Matter | Black Hole Masses & AGN Structure | Expansion History of the universe |
|---|---|---|---|---|
| **Key Science Questions** | | Is dark matter cold? What is the mass of the dm particle? | What are the masses and accretion disk sizes of Black Holes which fuel AGN? | What is Hubble's constant to sub-1% accuracy? |
| Science Requirements | | 0.1-0.8% time-delay precision for perturbation measurement | Microlensing events observed in multiple wavelength bands | 0.1-0.8% time-delay precision in full sample |
| | | Flux ratios between multiple images | Spatially-resolved R>1000 spectra over ~12x12arcsec | Multiple image separations, flux ratios |
| Measurement Requirements | Wavelength-Imag. | ≤300 to ~1100nm | ≤300 to ≥1100nm | ~300 to ~1100nm |
| | WavelengthSpec/y | ~400 to 800nm | 400 nm – 800 nm | -- |
| | Angular Resolution Imag. | 0.11arcsec at 650nm | 0.11arcsec at 650nm | 0.11arcsec at 650nm |
| | Spectral Resolution Spec/y | R ~ 2000 | R > 1000 | -- |
| | Signal To Noise | ≥100 (1% point-source photometry) | >10 per resolution element spec/y | ≥100 (1% point-source photometry) |
| | Cadence | Imaging: fast cadence (hours to days) for revisit observations | Spectra: slow cadence (days to weeks) for revisit observations | Imaging: fast cadence (hours to days) for revisit observations |
| Driving Mission Requirements | Primary Mirror Diameter | 1.5 m | >1.2 m | 1.5 m |
| | Orbit | L2 | L2 | L2 |
| | Mission duration | 3+ years | 5 years | 3+ years |
| | Pointing Accuracy | 30arcsec (3σ) | 30arcsec (3σ) | 30arcsec (3σ) |
| | Pointing Stability | 0.09arcsec over 600 sec (3σ) | 0.09arcsec over 600 sec (3σ) | 0.09arcsec over 600 sec (3σ) |
| | Pointing Knowledge | 15arcsec (3σ) | 15arcsec (3σ) | 15arcsec (3σ) |
| | Optical design | 6 channel dichroic splits | 6 channel dichroic splits | |
| | Thermal design | Passive control, T <~ 150K, stable | Passive control, T <~ 150K, stable | |
| | Field of View | 5.2x5.2 arcminutes | 5.2x5.2 arcminutes | |

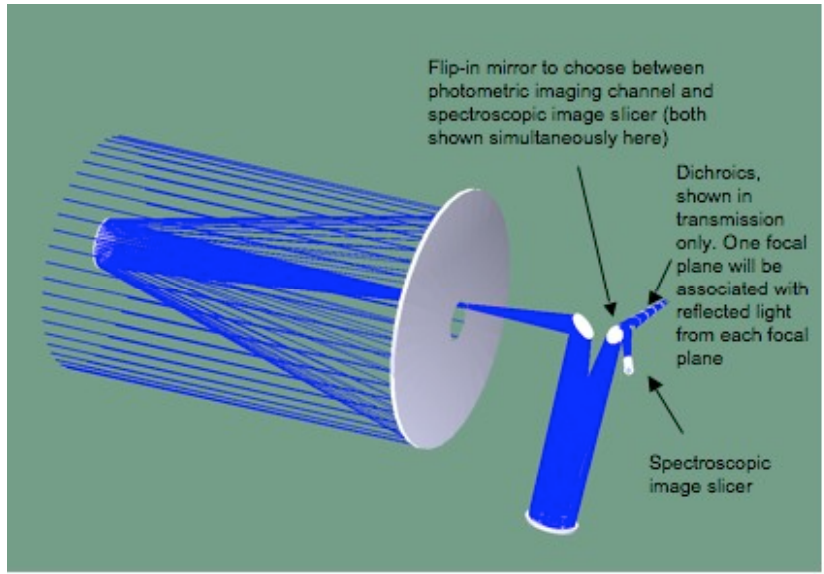

Figure 4. The OMEGA 1.5m telescope, showing both instrument channels. The default observing mode is with the multi-band photometric imager. Light is sent to the spectroscopic image slicer through a flip-in mirror.

### 3.2 Instruments

OMEGA has two instruments: a set of six focal planes for precision multi-wavelength photometry, and a second channel enabled by a flip-in mirror for an image slicer for spectroscopy with a single focal plane. OMEGA will operate in one mode at a time. The scientific and instrument design requirements flow-down can be summarized into three major requirements for the detectors in each focal plane. These include requirements for detectors that cover a broad spectral range (300 nm to 1100 nm minimum), with large dynamic range, high efficiency, high resolution, and excellent photometric stability (<1%).

#### 3.2.1 Photometric Channels

The OMEGA imager consists of a system of six cascading thin-film dichroics, each reflecting prescribed wavelength ranges to a separate dedicated detector, transmitting the remaining light bundle to the next step. By use of the dichroics, all bands image the same 5.2x5.2 arcmin field of view *simultaneously*, centered on each target. With the diffraction limit of this 1.5m aperture set at 650 nm, the 0.11 arcsec point spread function (PSF) is perfectly adapted for simultaneous high-precision photometry of the images in strong lenses which have 0.1-2.0 arcsec separations, making for a system that is many times more efficient than an imager with a filter wheel.

Table 2 OMEGA optical parameters

| Survey Mode | Wavelength Range by Channel | Resolution |
|---|---|---|
| Photometric, designed to simultaneously image the same field in all channels. | *300 – 1100 nm*<br>Ch 1: 300-400 nm<br>Ch 2: 400-550 nm<br>Ch 3: 550-700 nm<br>Ch 4: 700-820 nm<br>Ch 5: 820-960 nm<br>Ch 6: 960-1100nm | Diffraction-limited at 650nm (0.11 arcsec) |
| Integral Field Image Slicer | *400 – 800nm* | $R = \lambda/d\lambda \sim$ 2000<br>0.1arcsec slices |

The OMEGA spectral range is baselined from 300 to 1100nm. There are advantages to a system that covers even shorter wavelengths (reaching ~200nm) and extending toward 1400nm, and this is currently under study. The detectors are required to have high efficiency, low noise, and a full-well capacity that are compatible with the noise floor and dynamic range requirements. These trade with the pixel size as smaller pixels enable higher resolution for a given detector size, but also limit the dynamic range. Pixel size is also a key design parameter for the optical system itself.

To improve reliability, reduce system complexity for electronics and testing, and reduce costs, the *same* CCD choice will be baselined for all focal planes. The baseline detectors are the Lawrence Berkeley National (LBNL) p-channel fully-depleted CCDs with 3.5K x 3.5K layout, and 10.5micron pixel pitch, processed with the delta-doping technique to improve the quantum efficiency response especially at shorter wavelengths. There are e2V detectors at higher technology readiness levels (TRLs) that may not have the full benefits of the LBNL CCDs (which are more "red-sensitive"), which we will address below, and also continue to study.

### 3.2.2 Image Slicer

A simple movable mirror redirects the full beam from the photometric imager to the spectroscopic channel, a 6x6 arcsec field-of-view integral field unit image slicer (IFS). The IFS has a spectroscopic resolution of R~2000, spanning 400-800nm, obtaining spectroscopy of its field (covering both the lens galaxy and the lensed AGN images) in 0.1arcsec-width slices. The full design of the IFS is currently work in progress.

### 3.2.3 Pointing Control

The pointing requirements are shown in Table 1. A pointing accuracy of 30arcsec and onboard pointing knowledge of 15arcsec are achievable with the SA-200HP. A pointing stability of 0.09arcsec over the 600s integration time can be achieved with a fine guidance sensor system (FGS). The FGS may have a dedicated focal plane, or use pixels in at one of the focal planes for calibration of the telescope boresight pointing relative to the spacecraft attitude. Some ACS hardware (e.g., star trackers) may also need to be upgraded.

Four reaction wheels are used for fine pointing control. Wheel speeds are maintained within a limited range (e.g. 200 to 1500 RPM) to avoid zero-crossings and internal wheel resonances. Thrusters are used for delta-V maneuvers and unloading excess angular momentum. Sensor and actuator placement need to be considered to minimize interactions between the control system and the flexible structure, as well as potential contamination issues for the telescope if we extend the wavelength to operate below 300nm.

## 3.3 Technology Readiness

OMEGA plans to use high TRL components. The driving technological areas, for which a technology roadmap is being designed, include: optimized CCD detectors sensitive in the UV for precision photometry (TRL 4-5); and mirror coatings and dichroic designs to maximize bandwidth while minimizing risk (TRL 6 above 300 nm, TRL 3 from 200-300nm). An attitude control system with a Fine Guidance Sensor (FGS) for OMEGA's required pointing stability (TRL 6+) using *Spitzer*, *Kepler*, and *JWST* heritage is considered an engineering challenge, but not a technological innovation.

### 3.3.1 Detector Technologies

OMEGA's baseline detectors are the LBNL CCDs, which are of moderate TRL (TRL 6 for the sensors, TRL 4-5 for the electronics/packaging as determined by the Beyond Einstein Program Assessment Committee report). These detectors may be delta-doped to produce reflection-limited response throughout the 200-1000 nm range. Delta-doping is a processing technique available from JPL for virtually any commercially fabricated silicon detectors, and that significantly improves the quantum efficiency of these detectors in the near-ultraviolet. Though the LBNL CCDs have been designed for increased performance near 1 micron, the longest wavelength focal plane (960-1100 nm) could be covered with HgCdTe Hawaii 2RG chips available from Teledyne (formerly Rockwell Scientific Company), instead. Additional alternative detectors are available commercially from the vendor e2v, including 2K x 4K 13.5 micron pixel pitch, thinned deep depletion CCDs, and being at TRL 6+, are worth further consideration. Full depletion is better for the resolution and the PSF, especially at the shorter-wavelength channels. Back-illuminated detectors provide the highest achievable quantum efficiency and the best overall performance.

An extension of the wavelength regime to 1400nm, which is under consideration, would require a second set of detectors beyond the baseline LBNL (or e2v) CCDs. HgCdTe detectors are the most advanced technologically. Smaller formats (≤1Kx1K) are at higher TRL and are available from Teledyne. The *James Webb* Space Telescope has tested 2K x 2K

HgCdTe detectors from Teledyne to TRL 6, and the SNAP collaboration is working to qualify commercially fabricated HgCdTe detectors for flight that would be appropriate for OMEGA. The pixel size of HgCdTe detectors is larger, ~18 microns, but still samples the PSF well enough at the longer wavelengths.

### 3.3.2 Mirror Coatings and Dichroic Design

*Mirrors*: The mirror materials that can be considered for all the mirrors in the system are protected silver and protected aluminum. Protected aluminum has high TRL and can capture the full spectrum from 200 nm to 1100 nm. However its reflectivity is low and it also presents a characteristic dip in reflectivity around 700 to 900 nm. With multiple reflections, the throughput suffers significantly. Secondly, when non-normal incidence is encountered, aluminum introduces undesirable polarization splitting. Silver on the other hand presents a reflectivity roll off below about 400 nm and cannot capture the near-ultraviolet. One compromise solution is to apply aluminum for the large primary mirror (to reduce complexity in manufacturing) and to apply multilayered UV enhanced silver for the rest of the smaller mirrors. A trade-off between UV capture and throughput loss is under study.

*Dichroic filters*: Technology for the design and manufacture of dichroic filters is relatively mature (TRL6+ above 300nm); however, the requirement to capture the full spectrum in multiple channels presents some challenges. Below 300nm, dichroic material properties are poorly known due to the proprietary nature of the applicable materials and processes (TRL 3). OMEGA's baseline short wavelength cutoff is 300 nm and at high TRL; nonetheless, it is worthwhile continuing to consider plans for how a 200 nm channel may be implemented. Figure 4 shows a representative throughput performance of baseline dichroic channels assuming a flat spectrum input to the system of dichroic filters. The sharpness of the bandpass edges, stop band attenuation and ripples in the pass bands can be tailored and optimized.

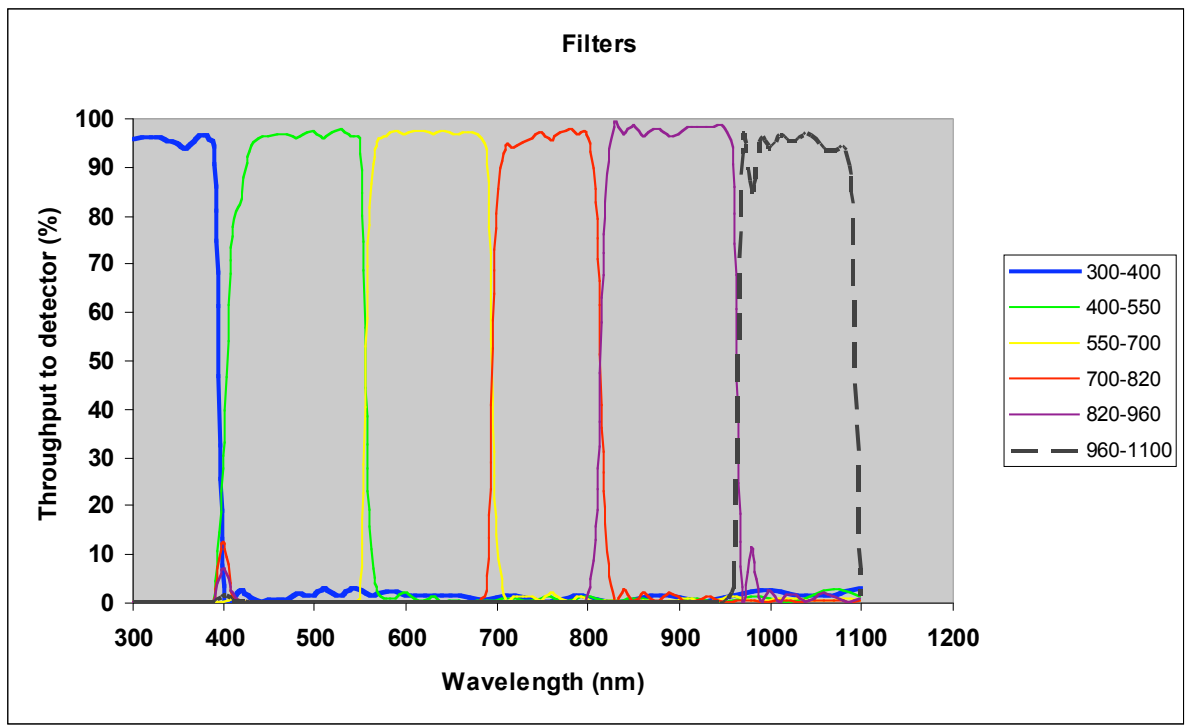

Figure 4. A representative design of dichroics tailored to specific wavelength ranges. The wavelength ranges given in the legend on the right correspond to the filter curves in order as shown, from left to right. The wavelength range is 300 – 1200 nm, and the full range of 0 – 100% throughput to detector in the ordinate axis. As it is possible to design thin-film dichroics with high throughput transmittance in specific wavelength ranges, and with extremely high reflectivity for all other wavelengths, the same light ray may be used for simultaneous multi-wavelength observations of the same field, removing the need for a filter-wheel (and the correspondingly large loss in on-target observing efficiency).

## 4. CONCLUSIONS

*OMEGA will lead to exquisitely fine views on the properties of dark matter and the nature of super-massive black holes, over cosmic time.* OMEGA will place extraordinary constraints on the properties of dark matter on astrophysical scales that are known to be extremely sensitive to the properties of the dark matter particle. These constraints will be possible at "snapshots" in space and time within individual lens galaxy halos, but also statistically as a function of cosmic time. Through the powerful combination of imaging and spectroscopy in the extended time domain, the detailed structure of a diverse sample of distant AGN will be probed from the inner-most edge of their accretion disks through the full broad-line region, and precise masses of their accreting black holes will be measured. By working with multiply-imaged strongly-magnified AGN, each with images too faint and too small a separation to tackle from the ground, OMEGA performs this experiment over a broad range of cosmic times and intrinsic AGN luminosities. The structure of AGN is tied to complicated and important fundamental physics about how matter behaves in the vicinity of extreme gravitational environments of super-massive black holes. These results will also shed a great deal of light on the connection between the co-evolution of AGN and galaxy formation and evolution. Finally, the cosmography possible with OMEGA (particularly with regard to an accurate Hubble's constant) is a complementary and nearly orthogonal technique to many others in vogue, and although it is achieved "for free" given the greater science drivers above, it will produce cosmography measurements which will play a strong supporting role in more focused expansion-history experiments.

## 5. ACKNOWLEDGMENTS

We are indebted to the expert contributions of Kunjithapatham Balasubramanian, Pantazis Mouroulis, Robert Kinsey, and Michael Hoenk at JPL. The work of PJM was supported by the TABASGO foundation in the form of a research fellowship, and in part by the U.S. Department of Energy, under contract number DE-AC02-76SF00515 at the Stanford Linear Accelerator Center. The research described in this paper was carried out at the Jet Propulsion Laboratory, California Institute of Technology, under a contract with the National Aeronautics and Space Administration.

## REFERENCES

[1] Keeton, CR, et al., "Differential Microlensing of the Continuum and Broad Emission Lines in SDSS J0924+0219, the Most Anomalous Lensed Quasar," ApJ, 639, 1 (2006)

[2] Peterson, BM & Bentz, MC, "Black Hole Masses from Reverberation Mapping," New Astronomy Reviews, 50, 796 (2006)

[3] Kochanek, CS, "Strong Gravitational Lensing", in Kochanek, Schneider, & Wambsganss, Part 2 of Gravitational Lensing: Strong, Weak, & Micro, Proceedings of the 33rd Saas-Fee Advanced Course, eds. G. Meylan, P. Jetzen, & P. North (Springer-Verlag: Berlin) (2004)

[4] Kochanek, CS et al., "Turning microlensing from a curiosity into a tool," in Statistical Challenges in Modern Astronomy IV, ed. GJ Babu & ED Feigelson (ASP Conference Series), 43 (2006)

[5] Mao, S & Schneider, P, "Evidence for substructure in lens galaxies?," MNRAS, 295, 587 (1998)

[6] Keeton, CR & Moustakas, LA, "A New Channel for Detecting Dark Matter Substructure in Galaxies by Time Delay Perturbations," ApJ, subm, 0 (2008)

[7] Zentner, A & Bullock, J, "Halo Substructure and the Power Spectrum," ApJ, 598, 49 (2003)

[8] Morgan, CW, et al., "The Quasar Accretion Disk Size – Black Hole Mass Relation," ApJL, subm, 0 (2007)

[9] Poindexter, S et al., "The Spatial Structure of an Accretion Disk," ApJ, 673, 34 (2008)

[10] Onken, CA, et al., "Supermassive Black Holes in Active Galactic Nuclei. II. Calibration of the Black Hole Mass-Velocity Dispersion Relationship for Active Galactic Nuclei," ApJ, 615, 645 (2004)

[11] Oguri, M, "Gravitational Lens Time Delays: A Statistical Assessment of Lens Model Dependencies and Implications for the Global Hubble Constant," ApJ, 660, 1 (2007).

[12] Olling, RP, "Accurate extragalactic distances and dark energy: anchoring the distance scale with rotational parallaxes," MNRAS, 378, 1385, (2007)